\documentclass{elsart}
\usepackage[intlimits]{amstex}
\usepackage{epsfig}

\begin{document}

\begin{frontmatter}
{\small\noindent UNITU-THEP-2/1998 \hfill FAU-TP3-98/1}

\vspace{5mm}

\title{Solving the Gluon Dyson--Schwinger Equation in the 
        Mandelstam Approximation} 
  \author[Tue]{A.~Hauck},
  \author[Erl]{L.~von Smekal},
  \author[Tue]{R.~Alkofer}
  \address[Tue]{Auf der Morgenstelle 14,
           Institut f\"ur Theoretische Physik,
           Universit\"at T\"ubingen,
           72076 T\"ubingen, Germany}
  \address[Erl]{Institut f\"ur Theoretische Physik III,
           Universit\"at Erlangen--N\"urnberg,
           Staudtstr.~7, 91058 Erlangen, Germany}

  \begin{abstract}

  Truncated Dyson--Schwinger equations represent finite subsets of the
  equations of motion for Green's functions. Solutions to these non--linear
  integral equations can account for non--perturbative correlations. We
  describe the solution to the Dyson--Schwinger equation for the gluon
  propagator of Landau gauge QCD in the Mandelstam approximation. This involves
  a combination of numerical and analytic methods: An asymptotic infrared
  expansion of the solution is calculated recursively. In the ultraviolet, the
  problem reduces to an analytically solvable differential equation. The
  iterative solution is then obtained numerically by matching it to the
  analytic results at appropriate points. Matching point independence is
  obtained for sufficiently wide ranges. The solution is used to extract a
  non--perturbative $\beta$--function. The scaling behavior is in good
  agreement with perturbative QCD. No further fixed point for positive values
  of the coupling is found which thus increases without bound in the infrared. 
  The non--perurbative result implies an infrared singular quark interaction
  relating the scale $\Lambda $ of the subtraction scheme to the string tension
  $\sigma$. 

  \vskip .2cm


  \end{abstract}
\end{frontmatter}

\newpage

{\Large PROGRAM SUMMARY}

\textit{Title of program:} mandelstam

\textit{Catalogue identifier:}

\textit{Program obtainable from:}
   CPC Program Library, Queen's University of Belfast, N.~Ireland

\textit{Computers:} Workstation DEC Alpha 500

\textit{Operating system under which the program has been tested:} UNIX

\textit{Programming language used:} Fortran 90

\textit{Memory required to execute with typical data:} 200 kB

\textit{No. of bits in a word:} 32

\textit{Peripherals used:} standard output, disk

\textit{No.~of lines in distributed program, including test data, etc.:} 347


\textit{Nature of physical problem:} \\
   An approach to describing non--pertubative correlations in field
   theories is to investigate their Dyson--Schwinger equations in suitable
   truncation schemes. Thereby one generally encounters non--linear integral
   equations which in many cases have infrared singular solutions imposing     
   stability problems in the numerical procedures.

\textit{Method of solution:} \\
   The infrared singular part of the integral equation is studied analytically
   and transformed into constraints. Infrared subleading conributions are
   then obtained in form of an asymptotic series. This expansion of the
   solution can be calculated recursively and, together with the ultraviolet
   behavior obtained from a differential equation, it makes a numerical
   solution tractable. 

\textit{Restrictions on the complexity of the problem:} \\
   So far all contributions from Faddeev--Popov ghosts are neglected.
   An extended model based on analogous techniques is presented elsewhere
   \cite{Hau97b,Sme97}.

\textit{Typical running time:} Approximately 10 sec.


{\Large LONG WRITE-UP}

\section{The physical problem}

\subsection{Introduction}

The knowledge of the infrared behavior of the running coupling 
$\alpha_S(\mu^2)$ in QCD is crucial for an understanding of confinement.
Should the theory have no further fixed point for $\alpha_S > 0$ then the
running coupling increases without bound in the infrared, and confinement
might be realized by an absence of the cluster decomposition property in QCD.

Due to the intrinsically non--perturbative nature of the problem, very
little is known about the strong coupling $\alpha_S$ in the infrared. This
is in contrast to its accurate knowledge in the ultraviolet, obtained via
the perturbative calculation of the Callan--Symanzik $\beta$ function. The
logarithmic decrease of $\alpha_S$ signals asymptotic freedom and is
verified experimentally in an impressive manner. 

Non--perturbative methods are required to study the strong coupling $\alpha_S$
in the infrared. One important framework for non--perturbative studies of QCD
is provided by the infinite tower of its Dyson--Schwinger equations. These
studies rely on specific truncation schemes of this tower. In the present paper
we focus on an approximation scheme for the gluon Dyson--Schwinger equation in
Landau gauge originally proposed by Mandelstam \cite{Man79}. The corresponding
equation will be referred to as Mandelstam's equation  in the following. Two
distinct approaches to its numerical solution are reported in the literature
\cite{Atk81,Bro89}. 

It was already pointed out by Mandelstam that for selfconsistent solutions
to exist, certain conditions on the behavior of the gluon propagator in the
infrared have to be met \cite{Man79}. An existence proof, a discussion of
the singularity structure and an asymptotic expansion in the infrared for
the solution to Mandelstam's original equation can be found in \cite{Atk81}.
Taking care of various violations of gauge invariance in the approximation,
one arrives at a slightly different equation \cite{Bro89}. We present an
approach to the numerical solution of both equations, which combines 
methods of refs.\ \cite{Atk81} and \cite{Bro89}. The general arguments that
follow apply to both versions of the equation. 

To remove the ultraviolet divergences in the equation for the gluon
self--energy, wave--function renormalization is required, thus introducing an
arbitrary scale $\mu$. We demonstrate, how the renormalized equation can be
cast in a renormalization group invariant form. We solve this equation
numerically and determine the running coupling from the solution via the
renormalization condition. We show that the product of the coupling and gluon
propagator, $g D_{\mu\nu}(k)$, does not acquire multiplicative renormalisation
in the Mandelstam approximation. This allows to identify a physical scale, the
string tension, from the infrared singular result for this quantity. 
Consequently, our non--perturbative and renormalisation group invariant results
yield a relation between the string tension and the only parameter in our
calculation, the QCD scale $\Lambda$. This relation is in reasonable agreement
with the respective phenomenological values. We obtain a non--perturbative
$\beta$ function and recover the scaling behavior of perturbative QCD at the
ultraviolet fixed point modulo small corrections due to the presence of
ghosts in Landau gauge. For positive values of the coupling no further fixed
point exists. The running coupling increases without bound in the infrared.

The situation changes dramatically when ghosts are included
\cite{Sme97',Atk97}. Then the running coupling approaches an infrared stable
fixed point. As the corresponding calculations are very involved, however, we
demonstrate the numerical method within the less complex case of solving
Mandelstam's equation in this paper first. A generalization to the solution of
the coupled gluon--ghost equations will be presented elsewhere \cite{Hau97'}.
Nevertheless, the central idea of how to treat anticipated infrared
singularities in non--linear integral equations is much more transparent and
easier developed  from the simplified equations discussed here. The present
results allow furthermore for a detailed comparison of the two schemes. In
particular,  this allows to asses the influence and importance of ghosts in
Landau gauge.

\subsection{Mandelstam's approximation}

Besides all elementary two--point functions, i.e., the quark, ghost and gluon
propagators, the Dyson--Schwinger equation for the gluon propagator also
involves the three-- and four--point vertex functions which obey their own
Dyson--Schwinger equations. These equations involve successive higher n--point
functions. Typical truncation schemes for this infinite tower rely on
neglecting all higher but three--point functions expressing the latter in
terms of propagators. This can be done with additional sources of information
like the Slavnov--Taylor identities, which are entailed by gauge
invariance. In the present study we neglect fermions considering a pure
Yang--Mills theory, which is believed to reflect characteristic features of
QCD. Studies by Brown and Pennington \cite{Bro88} indicate that the inclusion
of quarks results in a suppression of the infrared part of the gluon
propagator, however, no qualitative changes were observed. In the Mandelstam
approximation ghosts are also neglected, because their contribution was
anticipated to be small \cite{Man79,Bro89}. As a result, one obtains a
simplified equation for the inverse gluon propagator  in momentum
space\footnote{We work in Euclidean space with a positive--definite metric
$\delta^{\mu\nu} = {\rm diag}(1,1,1,1)$; color indices are suppressed.}, 
\begin{multline} 
  {D^{-1}}^{\mu\nu}(k)
    = {D_0^{-1}}^{\mu\nu}(k) + \frac{g_0^2 N_C}{32\pi^4}
        \int \!d^4\!q \:  \Gamma_0^{\mu\rho\alpha}(k,q-k,-q)  \\
      \times D^{\alpha\beta}(q) D^{\rho\sigma}(k-q)
        \Gamma^{\beta\sigma\nu}(q,k-q,-k)  \quad ,
  \label{DSE00}
\end{multline}
where $D_0$ and $\Gamma_0$ are the bare gluon propagator and the bare
three--gluon vertex, and $\Gamma$ is the fully dressed vertex (we will use
$N_c=3$ colors). This together with a particularly simple form for the
three--gluon vertex is the Mandelstam approximation \cite{Man79}. 

\begin{figure}[t]
  \centerline{\epsfig{file=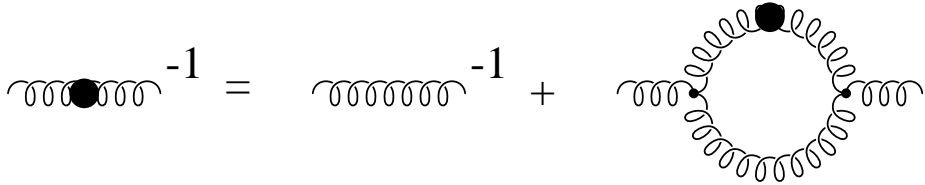,width=.9\linewidth}}
  \caption{ Diagrammatic representation of the gluon Dyson--Schwinger 
            equation in Mandelstam's approximation. }
  \label{fig:mandelstam}
\end{figure}

The form for the three--gluon vertex is motivated by its Slavnov-Taylor
identity, which neglecting all ghost contributions in the covariant gauge
reads:
\begin{equation}
  p^\mu \Gamma^{\mu\nu\rho}(p,q,k) =
    \frac{k^2}{Z(k^2)} \left( \delta^{\nu\rho} - \frac{k^\nu k^\rho}{k^2} \right)
    - \frac{q^2}{Z(q^2)} \left( \delta^{\nu\rho} - \frac{q^\nu q^\rho}{q^2} \right)
  \quad ,
  \label{eq:STI}
\end{equation}
where $p, q$ and $k$ are incoming momenta and $Z(k^2)$ is the invariant
function of the dressed gluon propagator, 
\begin{equation}
  D^{\mu\nu}(k^2) =
    \frac{Z(k^2)}{k^2} \left( \delta^{\mu\nu} - \frac{k^\mu k^\nu}{k^2} \right)
    + \xi \frac{k^\mu k^\nu}{k^4}  \quad .
  \label{glp}
\end{equation}  
The Slavnov--Taylor identity fixes the longitudinal part of the vertex.
It can be derived along the general procedure of ref. \cite{Bal80}. In
the simple version of Eq.\ (\ref{eq:STI}) without ghost contributions
the solution is: 
\begin{multline}
  \Gamma_L^{\mu\nu\rho}(p,q,k) =
    A_+(p^2,q^2) \delta^{\mu\nu} (p-q)^\rho +
    A_-(p^2,q^2) \delta^{\mu\nu} (p+q)^\rho   \\
    + C(p^2,q^2)( \delta^{\mu\nu} pq - p^\nu q^\mu) (p-q)^\rho +
    \text{cycl. permut.},
  \label{STIsol}
\end{multline}
with
\begin{equation}
  A_\pm (p^2,q^2) =
    \frac{1}{2} \left( \frac{1}{Z(p^2)} \pm \frac{1}{Z(q^2)} \right)
  \quad \text{and} \quad
  C(p^2,q^2) = 2 \frac{A_-(p^2,q^2)}{p^2-q^2} \; .
  \label{STIsol1}
\end{equation}
Assuming that $Z(q^2)$ is a slowly varying function one may write
\begin{equation}
  \Gamma_L^{\mu\nu\rho}(p,q,k)
    = A_+(p^2,q^2) \Gamma_0^{\mu\nu\rho}(p,q,k)  \; . \label{eq:MaAn}
\end{equation}
While this form for the full three--gluon vertex simplifies equation
(\ref{DSE00}) even more than the use of another bare vertex, $\Gamma
=\Gamma_0$, it is regarded superior to the latter since it accounts for
some of the dressing of the vertex as it results from the Slavnov--Taylor
identity. With $\Gamma = \Gamma_L$ from (\ref{eq:MaAn}) the Dyson--Schwinger
equation can now be cast in the simple form
\begin{eqnarray}  \label{eq:GluonDSE}
  {D^{-1}}^{\mu\nu}(k)
    &=& {D_0^{-1}}^{\mu\nu}(k) \,+\, \frac{g_0^2 N_C}{32\pi^4}
        \int \!d^4\!q \:  \Gamma_0^{\mu\rho\alpha}(k,q-k,-q) \nonumber \\
              && \hskip -.7cm \times D^{\alpha\beta}(q) D_0^{\rho\sigma}(k-q)
                \Gamma_0^{\beta\sigma\nu}(q,k-q,-k)  \quad . 
\end{eqnarray}
This equation is schematically depicted in Fig.~\ref{fig:mandelstam}.
In solving it an additional condition is implemented in the Mandelstam
approximation. Eq.\ (\ref{eq:STI}) and its solution
(\ref{STIsol},\ref{STIsol1}) for $p^2\to 0$ entail that
\begin{equation}
  \lim_{k^2 \to 0} \frac{k^2}{Z(k^2)} = 0,
  \qquad \text{or} \qquad
  \lim_{k^2 \to 0} D_{\!\mu\nu}^{-1}(k^2) = 0 \quad .
  \label{massless}
\end{equation}
While imposing this additional condition seems consistent with the other
assumptions in the Mandelstam approximation, it has to be emphasized that
concluding (\ref{massless}) as a result of (\ref{eq:STI}) relies on neglecting
all ghost contributions in covariant gauges. In fact, if ghosts are included
in resolving the Slavnov--Taylor identity for the 3--gluon vertex, condition
\eqref{massless} has to be modified \cite{Sme97'}.

In a manifestly gauge invariant formulation the Dyson--Schwinger equation for
the inverse gluon propagator in the covariant gauge would be transverse without
further adjustments. The longitudinal part of the gluon propagator does not
acquire dressing and cancels with the one of the bare propagator. This may be
violated due to the neglect of ghosts, the violation of Slavnov--Taylor
identities and also due to a regularization that does not preserve the residual
local invariance, i.e., the invariance under transformations generated by
harmonic gauge functions ($\partial^2 \Phi(x) = 0$). The latter is the case for
an $O(4)$ invariant Euclidean cutoff $\Lambda$, which we will use to regularize
Eq.\ (\ref{eq:GluonDSE}). In the present approximation, longitudinal terms can
be eliminated by  contracting Eq.\ (\ref{eq:GluonDSE}) with the transversal
projector,
\begin{equation}
  P^{\mu\nu}(k) = \delta^{\mu\nu} - \frac{k^\mu k^\nu}{k^2} \; .
\end{equation}
After performing the angular integrations the equation for the gluon
renormalization function $Z(k^2)$ in Landau gauge ($\xi =0$) then
becomes \cite{Man79} 
\begin{multline}
  \frac{1}{Z(k^2)} =
    1 + \frac{g_0^2}{16\pi^2} \int_0^{k^2} \frac{dq^2}{k^2}
              \left(   \frac{7}{8}\frac{q^4}{k^4}
                     - \frac{25}{4}\frac{q^2}{k^2}
                     - \frac{9}{2} \right) Z(q^2)  \\
      + \frac{g_0^2}{16\pi^2} \int_{k^2}^{\Lambda^2} \frac{dq^2}{k^2}
              \left(   \frac{7}{8}\frac{k^4}{q^4}
                     - \frac{25}{4}\frac{k^2}{q^2}
                     - \frac{9}{2} \right) Z(q^2) \; .
  \label{eq:Mandelstam}
\end{multline}
However, the above equation contains a quadratically ultraviolet divergent
term. This term violates the masslessness condition (\ref{massless}) and has
therefore been dismissed. As observed by Brown and Pennington \cite{Bro88},
in general, quadratic ultraviolet divergences can occur only in the part of
the inverse gluon propagator proportional to $\delta^{\mu\nu}$. Therefore,
that part cannot be unambiguously determined, it depends on the routing of
the momenta. This is due to the various violations of gauge invariance
mentioned above. The unambiguous term proportional to $k^\mu k^\nu$ can
be obtained by contracting (\ref{eq:GluonDSE}) with
\begin{equation}
  R^{\mu\nu}(k) = \delta^{\mu\nu} - 4 \frac{k^\mu k^\nu}{k^2} \quad .  
\end{equation}
In this case, upon angle integration instead of (\ref{eq:Mandelstam}) one
obtains \cite{Bro88} 
\begin{multline}
  \frac{1}{Z(k^2)} =
     1 + \frac{g_0^2}{16\pi^2}\int_0^{k^2} \frac{dq^2}{k^2} 
         \left(   \frac{7}{2}\frac{q^4}{k^4}
                - \frac{17}{2}\frac{q^2}{k^2}
                - \frac{9}{8} \right) Z(q^2)  \\
        + \frac{g_0^2}{16\pi^2} \int_{k^2}^{\Lambda^2}\frac{dq^2}{k^2}
          \left(   \frac{7}{8}\frac{k^4}{q^4}
                 - 7\frac{k^2}{q^2} \right) Z(q^2)  .
  \label{eq:Brown}
\end{multline}
It is clear that the assumptions made so far are drastic, and it is questionable
how much of the original theory survived.Furthermore, inclusion of ghosts
changes the conclusions on the infrared behavior of gluon propagator and the
running coupling even qualitatively \cite{Sme97'}. For the sake of a detailed
understanding of what exactly can be attributed to ghosts, it is nevertheless
necessary to thoroughly study the implications of the Mandelstam approximation,
most remarkably on the running coupling, in order to compare to conclusions of
the present and earlier work on the Dyson--Schwinger equations of Landau gauge
QCD. Besides this, the major numerical concepts used in the present paper,
entirely devoted to the Mandelstam approximation, will apply to the coupled
system of ghosts and gluons in Landau gauge with little changes.

One might think that the particular problem problem with ghosts can be avoided
using gauges such as the axial gauge, in which there are no ghosts in the first
place. As for studies of the gluon Dyson--Schwinger equation in axial gauge
\cite{BBZ81,Scho82,CuRo91} it is important to note that these rely on an
\textit{ad hoc} assumption on the tensor structure of the gluon propagator. The
occurrence of a second independently possible term, which happens to vanish in
perturbation theory, has been disregarded so far \cite{Bue95}. In fact, if the
complete tensor structure of the gluon propagator in axial gauge is taken into
account properly, one arrives at a coupled system of equations which is of
similar complexity as the ghost--gluon system in the Landau gauge
\cite{MPpriv}. Progress is desirable in axial gauge as well, of course. A
necessary prerequisite for this, however, is a proper treatment of the spurious
infrared divergences which are well known to be present in axial gauge due to
zero modes. This can be achieved by either introducing redundant degrees of
freedom, i.e.\ ghosts, also in this gauge (by which it obviously looses its
particular advantage) or by using a modified axial gauge \cite{Len94}, which is
specially designed to account for those zero modes. Eventually, progress in
more than one gauge will be the only reliable way to asses the influence of
spurious gauge dependencies.

\subsection{Renormalization}

The logarithmic ultraviolet divergences in equations (\ref{eq:Mandelstam})
and (\ref{eq:Brown}) can be removed by multiplicative renormalization.
Introducing the renormalized gluon propagator $D_R$ and the renormalized
coupling $g$ the renormalization constants $Z_3$ and $Z_g$ are defined by 
$D = Z_3 D_R$ and  $g_0 = Z_g g$ respectively. From the Dyson--Schwinger
equation for the gluon propagator in Mandelstam approximation
(\ref{eq:GluonDSE}), renormalized this way, we obtain
\begin{equation}  \label{eq:gluon-ren0}
  Z_R(k^2) = \left[ Z_3 +   Z_3^2 Z_g^2 \, \frac{g^2}{16\pi^2}\,
              I_{Z_R}(k^2) \right]^{-1} \; .
\end{equation}
Here, $I_{Z_R} := I[Z_R]$ is a functional of $Z_R(k^2)$; for Eq.\
(\ref{eq:Brown}) $I_{Z_R}(k^2)$ is given by  
\begin{multline}
  I_{Z_R}(k^2) =
    \frac{1}{k^2} \int_0^{k^2} dq^2 \left( \frac{7}{2}\frac{q^4}{k^4}
    - \frac{17}{2}\frac{q^2}{k^2} - \frac{9}{8} \right) Z_R(q^2)  \\
    + \frac{1}{k^2} \int_{k^2}^{\Lambda^2} dq^2
    \left( \frac{7}{8}\frac{k^4}{q^4} - 7\frac{k^2}{q^2} \right) Z_R(q^2)\;  .
  \label{eq:gluon-ren1}
\end{multline}
An analogous form follows for Mandelstam's original equation. After the
subtraction of the ultraviolet divergences, we use the limit $\Lambda \to 
\infty$ for the cutoff $\Lambda$ implicitly. Confessing that we are dealing
with renormalized quantities in the following, we dismiss the subscript $R$
again. Below we will prove the identity $Z_g Z_3 = 1$ for the Mandelstam
approximation. Furthermore, we adopt a momentum subtraction scheme requiring
the gluon self--energy to vanish at the renormalization point $\mu$, i.e., 
\begin{equation}
  Z(\mu^2) = 1
  \qquad \Rightarrow \qquad
  Z_3 = 1 - \frac{g^2}{(4\pi)^2} \, I_Z (\mu^2) \; .
  \label{Ren-Bed}
\end{equation}
With this and $\alpha_S(\mu^2) = g^2(\mu^2)/4\pi$ Eq.\ \eqref{eq:gluon-ren0}
reads, 
\begin{equation}
\label{eq:gluon-ren}
  Z(k^2) = \left[ 1+\frac{\alpha_S(\mu^2)}{4\pi}
              \left[ I_Z(k^2)-I_Z(\mu^2) \right] \right]^{-1} \; .
\end{equation}
To determine the behavior of $Z(k^2)$ for $k^2 \to 0$ we first notice that
\begin{enumerate}
\item for a constant $Z(k^2)$ the last term in (\ref{eq:gluon-ren1}) yields
      an infrared logarithm. Therefore, a solution to (\ref{eq:gluon-ren})
      cannot approach a constant as $k^2 \to 0$. Any constant term in the
      denominator on the r.h.s.\ of (\ref{eq:gluon-ren}) has to vanish. This
      is a first constraint for a possible solution.
\item Similarly $Z(k^2) \sim k^2$ as $k^2 \to 0$ is inconsistent
      since then the next to leading term yields a logarithmic contribution.
\item Now let us assume $Z(k^2) \sim 1/k^2$. This form yields exclusively
      terms that violate the condition (\ref{massless}). Such terms have to
      be subtracted. Since the kernels of all integrals are linear in $Z$,
      this is achieved by simply subtracting a corresponding contribution
      from $Z$ in the integrands. Thus defining $F(k^2)$ by
      \begin{equation}
        Z(k^2) =  \frac{b}{k^2} + F(k^2) 
      \end{equation}
      with some constant $b$, we further explore Ans\"atze for $F$, which is
      the remainder of $Z$ in the integrals.
\end{enumerate}

To isolate the infrared singular term $b/k^2$ we rewrite equation
(\ref{eq:gluon-ren}) as:
\begin{equation}
  Z(k^2) = \frac{b}{k^2} + F(k^2) = \frac{1}{A + B k^2 + J(k^2)} \; ,
  \label{eq:13}
\end{equation}
where
\begin{align}
  A &= 1 - \frac{\alpha_S(\mu^2)}{4\pi}\left[ 7\int_0^{\Lambda^2} dq^2 \:
           \frac{F(q^2)}{q^2} + I_F(\mu^2) \right] , \label{eq:14}\\
  B &= \frac{7}{8}\frac{\alpha_S(\mu^2)}{4\pi} \int_0^\infty dq^2 \:
        \frac{F(q^2)}{q^4}\; , \label{eq:15}\\
  J(k^2) &=  \frac{\alpha_S(\mu^2)}{4\pi} \int_0^{k^2} dq^2  
  \biggl(  \frac{7}{2}\frac{q^4}{k^6} - \frac{17}{2}\frac{q^2}{k^4}
            - \frac{9}{8}\frac{1}{k^2}
    + 7 \frac{1}{q^2} - \frac{7}{8}\frac{k^2}{q^4}
        \biggr) F(q^2) \; .
  \label{eq:16} 
\end{align}
As already stated in (1) above, a possible solution will have to obey $A=0$.
We will verify below that $A=0$ is not an independent constraint but an
identity. In addition, we see that $b=1/B$ must be met, which will be
imposed as a constraint on a possible solution.
  
Introducing dimensionless variables and appropriately scaled gluon self--energy
functions $\tilde{Z}$ and $\tilde{F}$ by 
\begin{equation}
  x = \frac{k^2}{b} \sqrt{\frac{4\pi}{\alpha_S(\mu^2)}}
    =: \frac{k^2}{\lambda^2}
  \quad \hbox{and} \quad
  \tilde{Z}(x)
    = \sqrt{\frac{\alpha_S(\mu^2)}{4\pi}} \, Z(k^2)
    = \frac{1}{x} + \tilde{F}(x) \; ,
  \label{eq:scale}
\end{equation}
the equation under consideration can be rendered dimensionless and
scale--independent. From equation (\ref{eq:13}) we obtain
\begin{equation}
  \frac{x}{1 + x \tilde{F}(x)} =
     \tilde{A} + \tilde{B} x + \int_0^x \frac{dy}{x}
     \left( \frac{7}{2}\frac{y^2}{x^2} - \frac{17}{2}\frac{y}{x}- \frac{9}{8}
           + 7 \frac{x}{y} - \frac{7}{8}\frac{x^2}{y^2} \right) \tilde{F}(y) \; .
  \label{eq:prel}
\end{equation}
Here, 
\begin{equation}
  \tilde{B} = b B = \frac{7}{8} \int_0^\infty dy \: \frac{\tilde{F}(y)}{y^2} 
  \stackrel{!}{=} 1  \; ,
  \label{eq:constraint}
\end{equation}
from the constraint mentioned above. Since $\tilde{F}(x)$ vanishes for
$x\to 0$ as will be shown later, in this limit equation (\ref{eq:prel}) shows
explicitly that the $x$-independent constant 
\begin{equation}
  \tilde{A} = \frac{4\pi}{g} A = 0 \; .
\end{equation}
With (\ref{eq:constraint}) equation (\ref{eq:prel}) therefore yields
\begin{equation}
  \frac{x^2 \tilde{F}(x)}{1+ x \tilde{F}(x)} =
    -\int_0^x \frac{dy}{x} \left( \frac{7}{2}\frac{y^2}{x^2} -
    \frac{17}{2}\frac{y}{x} - \frac{9}{8} + 7\frac{x}{y} -
    \frac{7}{8}\frac{x^2}{y^2} \right) \tilde{F}(y) \; .
  \label{eq:final}
\end{equation}
Before we turn to the numerical solution of (\ref{eq:final}) with the
constraint (\ref{eq:constraint}), some remarks need to be made.

First note that the renormalization condition $Z(\mu^2) = 1$ implies that 
\begin{equation}
  \tilde{\alpha}_S(s) = 4\pi\, \tilde{Z}^2(s) \; ,
  \label{alphaZ2}
\end{equation}
with $s := \mu^2/\lambda^2$ and $ \tilde{\alpha}_S(s) := \alpha_S(\mu^2)$.
This is compatible with the definition of $\tilde A$ which yields
($L := \Lambda^2/\lambda^2$)
\begin{align}
  \tilde{\alpha}_S(s)
   &= 4\pi \left[ \tilde A + 
        7\int_0^{L} \! dy \:
          \frac{\tilde{F}(y)}{y} +
        I_{\tilde F}(s) 
      \right]^{-2}   \label{eq:alpha}  \\
   &= 4\pi \biggl[ \tilde A + \tilde{B} s + \frac{1}{s} \int_0^{s} \! dy
          \left(
             \frac{7}{2}\frac{y^2}{s^2}
             - \frac{17}{2}\frac{y}{s}
             - \frac{9}{8} + 7\frac{s}{y}
             - \frac{7}{8}\frac{s^2}{y^2}
          \right) \tilde{F}(y) 
      \biggr]^{-2} \! . \nonumber
\end{align}
The right hand side above is identical to the one of (\ref{eq:prel}) with $x
\leftrightarrow s$, and thus (\ref{eq:alpha}) to (\ref{alphaZ2}). The fact
that only $\tilde{A} = 0$ is consistent, i.e., the infrared behavior of
the solution, entails that the running coupling is singular in the infrared. 

Furthermore, we observe that a solution to the system of equations
(\ref{eq:final}) and (\ref{eq:constraint}) is a renormalization group invariant
function. The argument  for this is as follows: Assume that the product $g
D_{\mu\nu}(k)$ does not acquire multiplicative renormalization. This means that
$\tilde{Z} \propto g Z$ is a renormalization group invariant combination. For
solutions of the form as given in (\ref{eq:13}) we conclude that
$\sqrt{\alpha_S/(4\pi)} b =: \lambda^2$ is a renormalization group invariant of
mass dimension two. Therefore, the introduction of the dimensionless variable
$x = k^2/\lambda^2$ in (\ref{eq:scale}) does not introduce an implicit
dependence on the renormalization scale $\mu$ in the explicitly
scale--independent set of equations (\ref{eq:final},\ref{eq:constraint}). This
verifies the assumption, i.e., a solution $\tilde{Z}(x) = g Z(k^2)/(4\pi)$ of
(\ref{eq:final},\ref{eq:constraint}) is a renormalization group invariant. For
the renormalization constants this entails that $Z_g Z_3 = 1$ in the Mandelstam
approximation. This is to be compared to the Abelian approximation to QCD, in
which $Z_g Z_3^{1/2} = 1$ or, equivalently, $g^2 D_{\mu\nu}(k)$ is a
renormalization group invariant combination (one application of this is
ref.~\cite{sme91}). Closer to the present case is the renormalization group
improved one--loop result in QCD. In perturbative QCD the power $\delta$ of the
coupling in the invariant combination $g^\delta D_{\mu\nu}(k)$ depends on the
number of quark--flavors $N_f$. Two examples are $\delta = 13/11$ for $N_f = 0$
and  $\delta = 1$ for $N_f = 3$. Even though $N_f = 0$ in Mandelstam
approximation, it resembles the three flavor result at this point. However, in
contrast to perturbation theory the identity $Z_g Z_3 = 1 (\delta = 1)$ holds
for all momentum scales $\mu$ in Mandelstam approximation.  

We conclude the discussion of the renormalization with a comment on the choice
of the non--perturbative subtraction scheme (\ref{Ren-Bed}). Since $g Z(k^2)$
is a dimensionless renormalization group invariant in Mandelstam
approximation, it is a function of the running coupling $\bar g(t,g)$,
\begin{equation}
  g Z(k^2) = f( \bar{g}(t_k, g)) \; ,
  \quad
  t_k = \frac{1}{2} \ln k^2/\mu^2 \; .
  \label{eq:gbar}
\end{equation}
Asymptotically, the momentum subtraction scheme is defined by $f(x) \to x$ for
$x\to 0$. If the product $gZ(k^2)$ is to have a physical meaning, e.g., as
potential between static color sources, it should be independent under changes
$(g,\mu) \to (g' , \mu ')$ according to the renormalization group. Taken the
idea of renormalization group invariance literally for arbitrary scales $\mu'$,
\begin{equation} 
  g Z({\mu'}^2) \stackrel{!}{=}  g' = \bar g(\ln (\mu'/\mu) , g) \; .
\end{equation} 
Then, $f(x) \equiv x$, $\forall x$, and, accordingly, $Z(\mu^2) = 1$, $\forall
\mu$. This is the only physically sensible non--perturbative extension of the
momentum subtraction scheme in the present context. Note that (\ref{eq:gbar})
with $f(x) \equiv x$ is identical to (\ref{alphaZ2}) with $s \leftrightarrow
x$, the renormalization scheme is thus equivalent to defining the
non--perturbative running coupling by the product of the coupling constant and
the gluon renormalization function in the Mandelstam approximation.

\section{Numerical methods}

\subsection{Series expansion in the infrared and asymptotic behavior in the
            ultraviolett}

A straightforward implementation of \eqref{eq:final} fails due to delicate
cancelations in the infrared. Therefore the expansion in the the infrared has to
be pushed beyond leading order for $\tilde{Z}(x)$. Making the Ansatz
$\tilde{F}(x) = a x^{\gamma}$ one sees that no solution exists for
$\gamma \leq 1$. For $\gamma>1$ eqn.~\eqref{eq:final} yields
\begin{equation}
  x^{2+\gamma} = -C_{00}(\gamma) x^\gamma - a C_{00}(\gamma) x^{1+2\gamma}
\end{equation}
with
\begin{equation}
  C_{00}(\gamma) =
    \frac{7}{2(\gamma+3)} - \frac{17}{2(\gamma+2)} - \frac{9}{8(\gamma+1)}
    + \frac{7}{\gamma} - \frac{7}{8(\gamma-1)} \quad .
\end{equation}
Obviously the coefficient $C_{00}(\gamma)$ has to vanish in order to permit
a selfconsistent solution. This yields a cubic equation with a unique
solution for $\gamma>1$:
\begin{equation}
  \gamma_0 =
    \frac{2}{9} \sqrt{229} \cos \left( \frac{1}{3} \arccos \left( 
    -\frac{1099}{229\sqrt{229}} \right) \right) - \frac{13}{9}
    \approx 1.2705 \quad .
  \label{eq:gamma0}
\end{equation}
An analogous condition was already found by Mandelstam for his original
equation \cite{Man79}. It is interesting to note that the value of
$\gamma_0$ given here is numerically very close to the one obtained by
Mandelstam ($\gamma_0 = \sqrt{\frac{31}{6}} - 1 \approx 1.273$), even
though the coefficients in (\ref{eq:Brown}) and in (\ref{eq:Mandelstam})
are completely different.

Looking at the corresponding results given in \cite{Bro89} one observes
that their infrared leading terms for $F(k^2)$ do not obey the constraint
(\ref{eq:gamma0}) for $\gamma_0$. We verified numerically that the
expressions given in ref.~\cite{Bro89} reproduce themselves under the
integral equation in the momentum range given in the corresponding figure
in this reference (Fig.~3 in \cite{Bro89}). However, trying to extend the
numerical calculations further to the infrared along the lines of
\cite{Bro89} we observed that $Z(k^2)$ became singular at lower but finite
$k^2$. Due to this fact an iterative numerical solution proved impossible
without constraining the infrared behavior of $Z(k^2)$ to that according
to the discussion of the solution given above. Therefore, we conclude that
while the expressions for $Z(k^2)$ in \cite{Bro89} with the parameters
given there are a numerical approximation to the solution of
Eq.~(\ref{eq:gluon-ren}), they do not represent the exact analytic result
particularly for $k^2 \ll \Lambda_{QCD}^2$.

For the numerical solution of Eq.~\eqref{eq:final} it turns out to be necessary
to study the subleading infrared behaviour of $\tilde{F}(x)$ in more detail.
This is possible by an asymptotic expansion of $\tilde{F}(x)$ at small $x$
analogous to what was introduced by Atkinson et al.\ for Mandelstam's original
equation in ref.\ \cite{Atk81},
\begin{equation}
  \tilde{F}(x)
    = \sum_{m=0}^M \sum_{n=0}^N \:
         a_{mn} \, x^{\gamma_0 + n\gamma_0 + n + 2m} \quad .
  \label{eq:series}
\end{equation}
The coefficients $a_{mn}$ can be determined from Eq.~\eqref{eq:final}:  
\begin{align}
  a_{0n} &= 0  \qquad\qquad \text{for $n>0$} \\
  a_{m0} &= \frac{a_{m-1,0}}{C_{m0}}  \qquad \text{for $m>0$} \nonumber \\
  a_{mn} &= \frac{1}{C_{mn}} \biggl(
               a_{m-1,n} - \sum_{m'=0}^{m-1} \sum_{n'=0}^{n-1} \:
               a_{m'n'}a_{m-m',n-n'}C_{m-m',n-n'}
            \biggr) \, , \; m,n > 0 \nonumber 
\end{align}
with the matrix $C_{mn}$ given by
\begin{align}
  C_{mn} =& - \frac{ 7}{2(\gamma_0 + n\gamma_0 + n + 2m + 3)}
            + \frac{17}{2(\gamma_0 + n\gamma_0 + n + 2m + 2)} \nonumber \\
          & + \frac{ 9}{8(\gamma_0 + n\gamma_0 + n + 2m + 1)} 
            - \frac{ 7}{  \gamma_0 + n\gamma_0 + n + 2m}      \nonumber \\
          & + \frac{ 7}{8(\gamma_0 + n\gamma_0 + n + 2m - 1)} \quad .
\end{align}
The double series \eqref{eq:series} is thus fixed up to the coefficient
$a_{00}$ of the infrared leading term. With this at hand the infrared region
up to a matching point $x_0$ is treated analytically and the equation for
$\tilde{F}(x)$ becomes:
\begin{multline}  \label{eq:Gluon-DSE+}
  \frac{x^2\tilde{F}(x)}{1+x\tilde{F}(x)} =
      \sum_{m=0}^M \sum_{n=0}^N \Biggl[ 
     - \frac{7}{2(\gamma_0 + n\gamma_0 + n + 2m + 3)}
         \left( \frac{x_0}{x} \right)^3  \\
     + \frac{17}{2(\gamma_0 + n\gamma_0 + n + 2m + 2)}
         \left( \frac{x_0}{x} \right)^2
     + \frac{9}{8(\gamma_0 + n\gamma_0 + n + 2m + 1)}
         \left( \frac{x_0}{x} \right)  \\
     - \frac{7}{\gamma_0 + n\gamma_0 + n + 2m}
     + \frac{7}{8(\gamma_0 + n\gamma_0 + n + 2m - 1)}
         \left( \frac{x_0}{x} \right)^{-1}  \Biggr]
    a_{mn} x_0^{\gamma_0+n\gamma_0+n+2m}  \\
    - \int_{x_0}^x \frac{dy}{y}  \left( \frac{7}{2}\frac{y^3}{x^3}
    - \frac{17}{2}\frac{y^2}{x^2} - \frac{9}{8}\frac{y}{x} + 7
    - \frac{7}{8}\frac{x}{y} \right) \tilde{F}(y)  \quad .
\end{multline}
Similarly the infrared part of the constraint \eqref{eq:constraint} is
rewritten with the help of the asymptotic series. In the numerical evaluation
of the (ultraviolet finite) integral in Eq.\ \eqref{eq:constraint} large
momentum contributions are also truncated. Though the resulting error becomes
negligible for a sufficiently large ultraviolet cutoff $x_1$,  convergence is
improved by correcting this error analytically. This is possible because
$\tilde{F}(x)$ can be replaced by its leading perturbative from for $x > x_1$
and sufficiently large $x_1$, which allows to calculate  the finite correction
term for the integration from $x_1$ to infinity on the r.h.s.\ of Eq.\
\eqref{eq:constraint} analytically. To obtain this ultraviolet behaviour, the
l.h.s.\ of eqn.~(\ref{eq:final}) is expanded for $x\tilde{F}(x) \rightarrow
\infty$. The leading terms cancel each other due to the constraint
(\ref{eq:constraint}). The next to leading order terms upon differentiation
yield the differential equation
\begin{equation}
  x \tilde{F}'(x) = -7\tilde{F}^3(x) \; ,
\end{equation}
which is solved by
\begin{equation}
  \tilde{F}(x) = \frac{1}{\sqrt{14 \ln x}}
  \quad \text{for} \quad x\rightarrow\infty \; .
\end{equation}
Going to higher orders one starts by expanding \eqref{eq:final}, neglecting
terms of order $1/x$ and higher and making use of Eq.\ \eqref{eq:constraint}.
Substituting $t=\sqrt{\ln x}$ one arrives at
\begin{equation}
  \frac{d}{dt} \frac{1}{\tilde{F}(t)}
    = \sqrt{14} + \frac{dg(t)}{dt}
    = \frac{14\,t}{\sqrt{14}\,t + g(t)}
      + \frac{10}{3\sqrt{14}}\frac{1}{t^2} + \ldots \; ,
\end{equation}
where we defined the function $g(t)$ for convenience. Thus 
\begin{equation}
  \frac{dg(t)}{dt} =
      -\frac{g(t)}{t} + \frac{10}{3\sqrt{14}}\,\frac{1}{t^2}
      + \mathcal{O}\left( \frac{g^2(t)}{t^2} \right)
\end{equation}
which is solved by
\begin{equation}
  g(t) = -\frac{c\sqrt{14}}{t} + \frac{10}{3\sqrt{14}}\,\frac{\ln t}{t}
         + \mathcal{O}\left( \frac{1}{t^4} \right)
         + \mathcal{O}\left( \frac{1}{\text{e}^{x^2}} \right) \; .
\end{equation}
Resubstituting for $x$ finally gives
\begin{equation}
  \tilde{F}(x) = \frac{1}{\sqrt{\kappa_0 \ln x}}
   - \frac{(\kappa_1/\kappa_0) \ln\ln x - \kappa_0 c}{(\kappa_0 \ln x)^{3/2}}
 + \mathcal{O} \left( \frac{1}{(\ln x)^{5/2}} \right)
        + \mathcal{O} \left( \frac{1}{x} \right) \; ,
  \label{eq:asympt0}
\end{equation}
where $\kappa_0 = 14$ and $\kappa_1 = 70/3$. The integration constant $c$
is difficult to determine numerically because of the unknown contribution of
the terms neglected and the extremely large scales necessary for the leading
logarithms to dominate. We find roughly that $c \approx 1$, its exact value
is irrelevant, however. With $x = k^2/\lambda^2$ and $\tilde{Z}(x) = g
Z(k^2)/(4\pi)$ the leading term in (\ref{eq:asympt0}) is equivalent to  
\begin{equation}
  Z(k^2) \to
     \frac{1}{\sqrt{\kappa_0/(4\pi)^2 \, g^2 \ln(k^2/\lambda^2 )}}
  \qquad
  \text{for $k^2 \to \infty$} \; .
\end{equation}
This resembles the renormalization group improved perturbative result,
\begin{equation}
  Z_{pt}(k^2) = \left( \beta_0 \, g^2 \ln k^2/\Lambda^2 \right)^{-\delta/2}
 \quad ,
 \label{pert-GLP}
\end{equation}
and allows for the following identifications:
\begin{itemize}
\item The Callan--Symanzik coefficient $\beta_0$ is given by $\beta_0 =
      \kappa_0/(4\pi)^2$, and the above value for $\kappa_0$ has to be
      compared to its perturbative result $\kappa_0 = 11 - (2/3) N_f$.
\item Since we use a momentum subtraction scheme, we identify the QCD scale
      to be used as $\Lambda_{\text{MOM}} = \lambda  $.
\item The anomalous dimension of the gluon field $\gamma_A (g)$ in
      perturbation theory is given by $\gamma_A (g) = \gamma_A^0 g^2$
      with $\gamma_A^0 = (13/2 - (2/3) N_f)/(4\pi)^2$, and the exponent
      in the r.h.s.\ of (\ref{pert-GLP}) is $\delta/2 = \gamma_A^0 /\beta_0$.
      The fact that $\delta = 1$ for the asymptotic behavior in the
      Mandelstam approximation confirms that $Z_g Z_3 = 1$, which is
      equivalent to $\gamma_A^0 = \beta_0 /2$. We thus obtain
      $\gamma_A^0 = 7/(4\pi)^2$.
\end{itemize}
The values for the scaling coefficients $\beta_0 $ and $\gamma_A^0 $
obtained from the Mandelstam approximation are reasonably close to their
perturbative values for $N_f =0$. The respective values for Mandelstam's
original equation (\ref{eq:Mandelstam}) are $\beta_0 = (25/2)/(4\pi)^2$ and
$\gamma_A^0 = (25/4)/(4\pi)^2$.

With the asymptotic infrared as well as ultraviolet contributions integrated
analytically, the constraint \eqref{eq:constraint} can now be written,   
\begin{multline}
  1 = \frac{7}{8} \Biggl[
  \sum_{m=0}^M \sum_{n=0}^N a_{mn}
  \frac{x_0^{\gamma_0+n\gamma_0+n+2m-1}}{\gamma_0+n\gamma_0+n+2m-1}  \\
  + \int_{x_0}^{x_1} dy \frac{\tilde{F}(y)}{y^2}
  + \sqrt{\frac{\pi}{14}} \operatorname{erfc} \sqrt{\ln x_1}  \Biggr] \; .
  \label{eq:constraint+}
\end{multline}

The integrals in eqn.~\eqref{eq:Gluon-DSE+} and \eqref{eq:constraint+} are
calculated with a Simpson integration routine of fourth order. The meshpoints
have been chosen equidistant on a logarithmic scale, i.e.\ we have substituted
\begin{equation}
  \int dy \longrightarrow \int du \, y
  \quad \text{with} \quad u = \ln y \quad .
\end{equation}
Note that choosing an equidistant mesh in $y$ will not result in a convergent
iterative process.

\begin{figure}
  \centerline{\epsfig{file=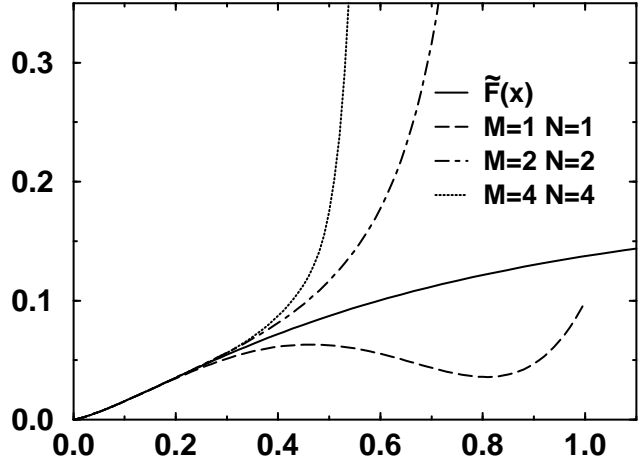,width=0.6\linewidth}}
  \caption{ The numerical solution $\tilde{F}(x)$ together with
            several asymptotic series of different order. }
  \label{fig:series}
\end{figure}

Equation \eqref{eq:Gluon-DSE+} can now be solved iteratively. Starting with 
a trial function for $\tilde{F}(x)$ the r.h.s.\ of (\ref{eq:Gluon-DSE+}) is
evaluated and, in order to account for the constraint, the result is rescaled
appropriately before it is used in the next iteration. To leading order the
contribution from the infrared expansion is proportional to $a_{00}$.
It is therefore sufficient to rescale $a_{00}$ linearly. This way, the
coefficient $a_{00}$ is determined numerically from the iteration process. 
For Eq.~(\ref{eq:Brown}) we obtain $a_{00} = 0.29446...$, and for
(\ref{eq:Mandelstam}) its value is $a_{00} = 0.29421...$. The infrared
matching point $x_0$ is chosen in a region around $0.2$. For this value it
proves sufficient to truncate the series (\ref{eq:series}) at $M=N=4$ (see
Fig.~\ref{fig:series}). For lower orders of the series we observe small bumps
at the matching point. Changing the matching point in a regime from 0.15 to
0.25 has no influence on the numerical solution. The lower limit is dictated
by convergence of the iterative process; the upper limit by a reasonable
accuracy of the asymptotic series. 

\begin{figure}
  \centerline{\epsfig{file=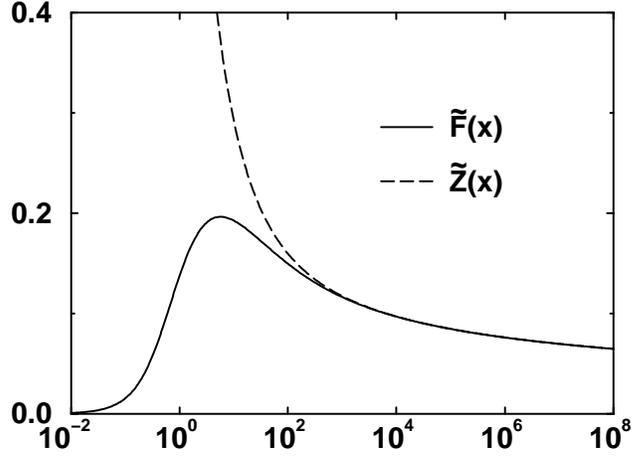,width=0.6\linewidth}}
  \caption{ The numerical solution $\tilde{F}(x)$ and for comparison
            $\tilde{Z}(x) = 1/x + \tilde{F}(x)$. }
  \label{fig:gluon}
\end{figure}

Using the numerical solution for $Z$, see Eq.~(\ref{eq:Brown}), the
renormalization group invariant product $gZ$ as a function of
$k^2/\Lambda_{\text{MOM}}^2$ is shown in Fig.~\ref{fig:gluon}. The solution
to Mandelstam's original equation (\ref{eq:Mandelstam}) is almost
indistinguishable on this scale.


\subsection{The running strong coupling}

\begin{figure}
  \epsfig{file=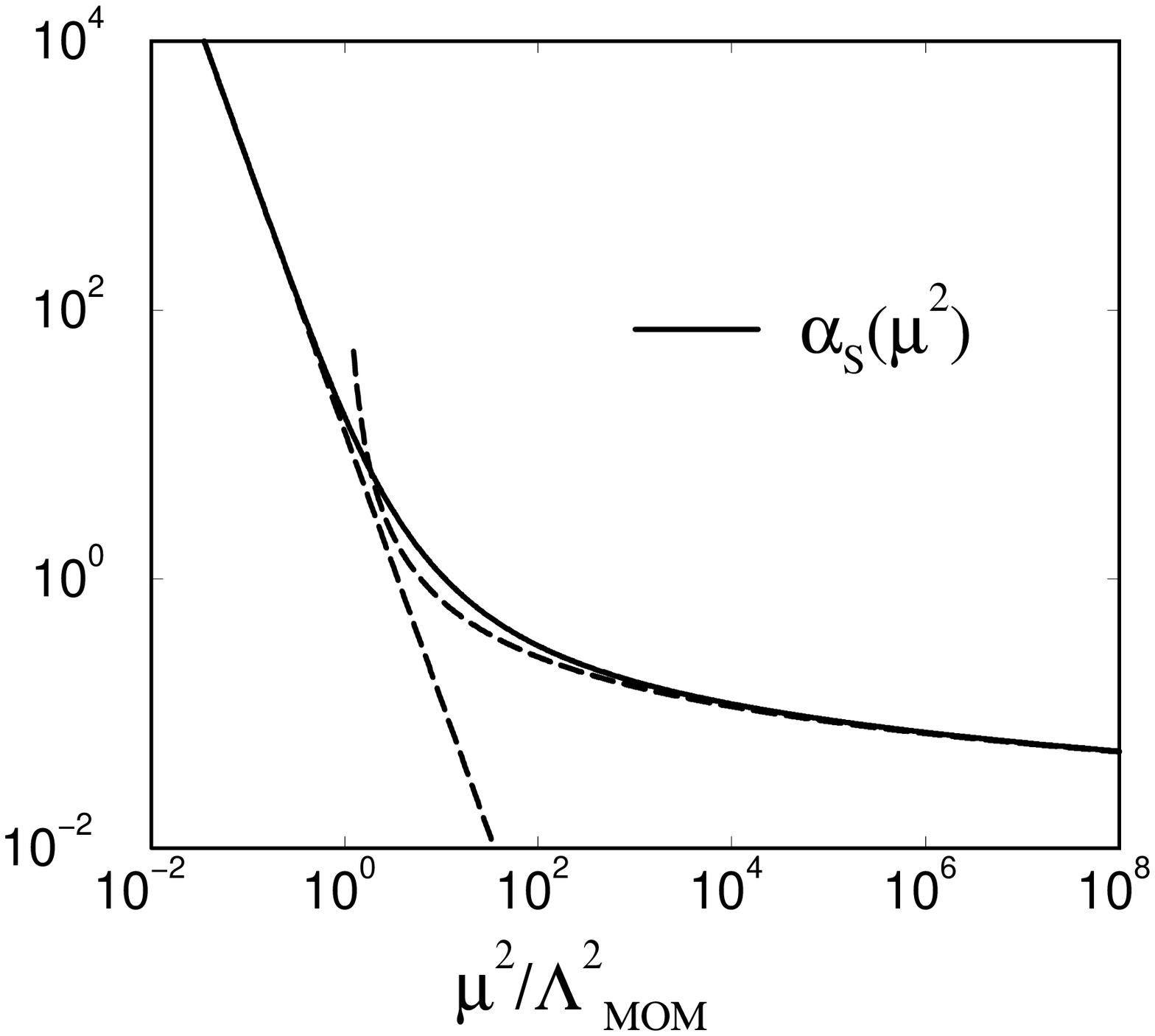,width=.45\linewidth}
  \hfill
  \epsfig{file=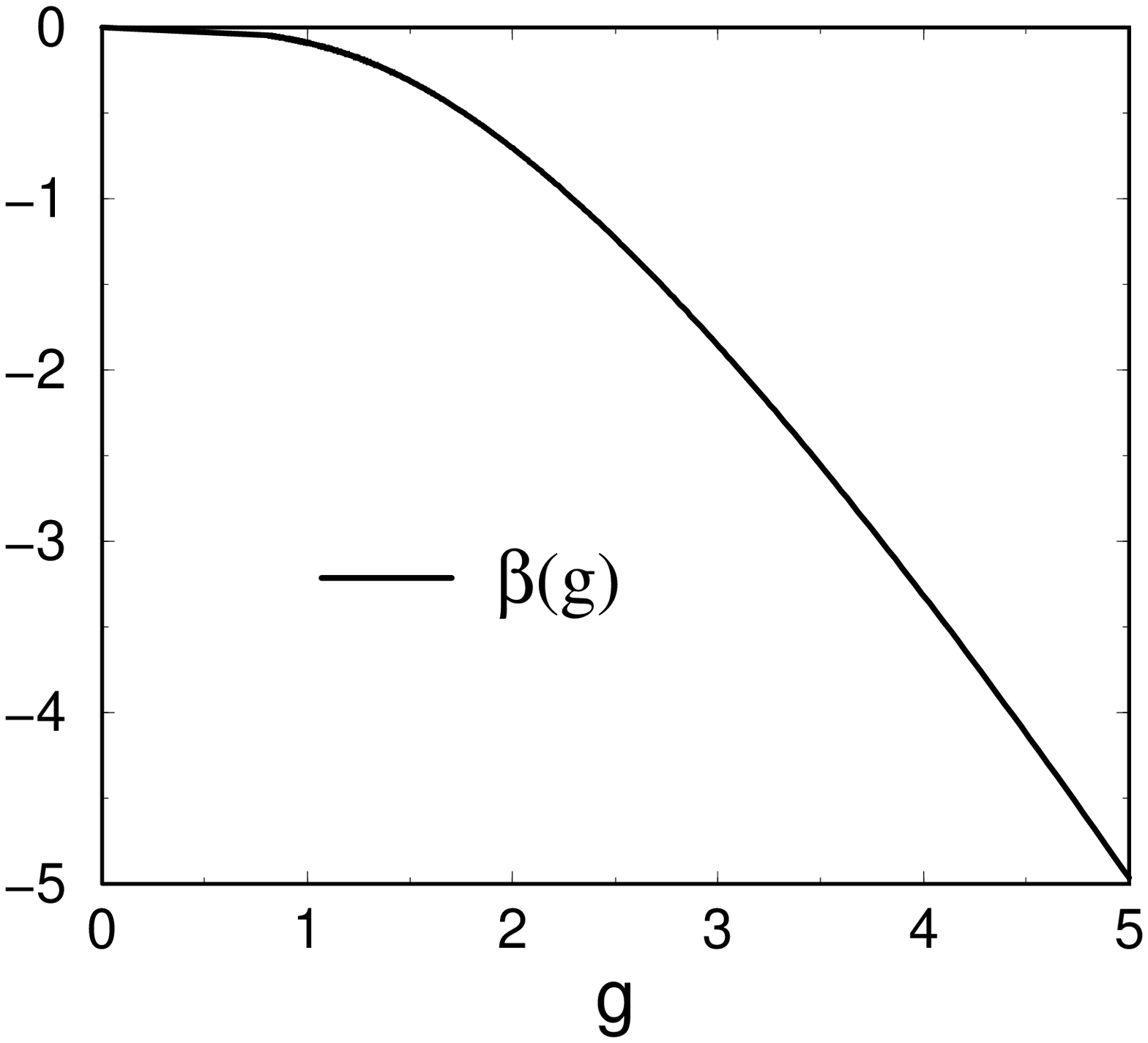,width=.45\linewidth}
  \caption{ The strong coupling $\alpha_S(\mu^2)$ with $\mu^2$ in units of
            the QCD scale $\Lambda_{\text{MOM}}^2$ and its asymptotic forms
            according to (\protect\ref{eq:alpha-0}) and
            (\protect\ref{eq:alpha-infty}) with $c=1$ (left); and the 
            non--perturbative $\beta$ function of the Mandelstam approximation
            (right). }  
  \label{fig:alpha}
\end{figure}

With the solution $\tilde Z(x)$ the running coupling $\alpha_S(\mu^2) = \tilde
\alpha_S(s)$ with  $s = \mu^2/\Lambda_{\text{MOM}}^2$ follows for the
subtraction scheme (\ref{Ren-Bed}) from Eq.~(\ref{alphaZ2}). The result
together with its analytically extracted asymptotic forms is shown in the left
graph of Fig.~\ref{fig:alpha}. In the infrared we have,
\begin{equation}
  \tilde\alpha_S(s) \to 4\pi \frac{1}{s^2}
  \quad \text{or} \quad 
  g(\mu^2) \to 4\pi \frac{\Lambda_{\text{MOM}}^2}{\mu^2} 
  \quad \text{for $\mu^2 \to 0$ .}
  \label{eq:alpha-0}
\end{equation}
In the ultraviolet one obtains from Eq.~(\ref{eq:asympt0}) the leading
logarithmic behavior,
\begin{equation}
\alpha_S(\mu^2) = \tilde{\alpha}_S(s) 
    = \frac{4\pi}{\kappa_0 \ln s}
     \left( 1 - \frac{2\kappa_1}{\kappa_0^2} \frac{\ln\ln s}{\ln s}
        + \frac{2c}{\ln s} \right)
   + \mathcal{O} \left( \frac{1}{(\ln s)^{3}} \right)
   + \mathcal{O} \left( \frac{1}{s} \right) \; .
  \label{eq:alpha-infty} 
\end{equation} 
This form resembles the two--loop perturbative result if the integration
constant $c$ is set to zero. The value of $\kappa_1 = 70/3$ compares to
$\kappa_1 = 51 - (19/3) N_f$ in perturbation theory \cite{PDG96}. The
Callan--Symanzik $\beta$ function, $\beta(g) = \mu (dg/d\mu )$, follows readily
for \textit{all} positive values of the coupling. The numerical result is shown
in the right graph of Fig.~\ref{fig:alpha}. Its limits are
\begin{equation}
  \beta(g) \to -\beta_0 g^3 \quad \text{for} \quad g \to 0
  \quad \text{and} \quad
  \beta (g) \to -2g \quad \text{for} \quad g \to \infty  \; .
\end{equation}
An alternative way to obtain the $\beta$ function follows directly from the
coupling renormalization $g_0 = Z_g g$. Because of $Z_g Z_3 = 1$ the gluon
field renormalization (\ref{Ren-Bed}) is equivalent to a renormalization
condition that relates the  $\mu$--independent bare coupling to the
renormalized one, 
\begin{equation}
  \frac{4\pi}{g_0}
     = 4\pi \frac{Z_3}{g}
     = \frac{4\pi}{g} - I_{\tilde Z}(\mu^2)  \; .
\end{equation}
This determines the $\beta$ function of the Mandelstam approximation with a
momentum subtraction scheme (fixing $Z(\mu^2) $ to a $\mu$ independent
constant as in (\ref{Ren-Bed})). Since $Z_g Z_3 = 1$ the anomalous dimension
of the gluon $\gamma_A(g)$ is,
\begin{equation}
  \gamma_A(g) =  -\frac{1}{2}\frac{\mu}{Z_3}\frac{dZ_3}{d\mu}
              =  -\frac{\beta(g)}{2g} \; .
\end{equation}
From the infrared behavior of the renormalization group invariant product
$gD_{\mu\nu}(k)$ we may deduce a linear rising potential between static color
sources for large distances. This allows us to relate the coefficient of the
$1/k^4$ term in the interaction to the string tension $\sigma$, 
\begin{equation} 
  \frac{gZ(k^2)}{k^2} \to 8\pi \frac{\sigma}{k^4}
  \quad \text{for} \quad k^2 \to 0 \; .
\end{equation} 
With the identification $\Lambda_{\text{MOM}}^2 = \lambda^2 $ (from the
ultraviolet) and the infrared behavior of the solution,
\begin{equation} 
  gZ(k^2) \to 4\pi \frac{\lambda^2}{k^2} \; , 
\end{equation}
the non--perturbative result for the gluon propagator relates the string
tension to the scale $\Lambda_{\text{MOM}}$, yielding  
\begin{equation}
  2\sigma  = \Lambda_{\text{MOM}}^2
\end{equation}  
in the Mandelstam approximation. The string tension can be fixed from
quarkonia potentials and Regge phenomenology \cite{Eic80,Buc81}. The result
is a value of about $\sigma = 0.18 \text{GeV}^2$. This has also been
confirmed in lattice calculations \cite{Din90}. Here, it corresponds to a
scale $\Lambda_{\text{MOM}}$ of 600 MeV. 

\section{Conclusions}

In this paper we have derived the infrared behavior of the strong coupling
constant from Mandelstam's approximation \cite{Man79} to the gluon
Dyson--Schwinger equation. This approximation results in a gluon propagator
diverging like $\sigma/k^4$ for $k^2\to 0$. This highly infrared singular
behavior of the gluon propagator generates a linearly rising potential between
static color sources. The corresponding string tension $\sigma$ can be used to
fix the scale in the resulting gluon self--energy. In particular, the
non--perturbative solution allows to relate the string tension to the QCD
scale $\Lambda_{\text{MOM}}$. This relation is in reasonable agreement with the
phenomenological values of the respective physical constants. Note that
dimensional transmutation occurs here: The coefficient $\lambda^2 = 2\sigma$ of
the infrared singular term is a dimensionful physical quantity, whereas all
input is dimensionless except for the renormalization scale $\mu$.

We have shown that the product of the coupling and the gluon propagator,
$gD_{\mu\nu}(k)$, does not acquire multiplicative renormalization in Mandelstam
approximation ($Z_g Z_3 = 1$). We calculated the gluon self--energy in a
renormalization group invariant fashion. Besides the running coupling
$\alpha_S(\mu^2)$ we obtained the non--perturbative Callan--Symanzik $\beta$
function and the anomalous dimension of the gluon field for the Mandelstam
approximation. In the ultraviolet the leading logarithmic decrease of the gluon
renormalization function allows to identify the scaling coefficients $\beta_0$
and $\gamma_A^0$ at the fix point. The scaling coefficients we obtain are in
reasonable agreement with the perturbative results. In addition, we have shown
that there is no further fix point for positive values of the coupling in the
Mandelstam approximation. The resulting strong coupling constant increases
without bound in the infrared.

We have demonstrated how to solve an integral equation with infrared
singlular solution. This combines numerical and analytic methods based on
asymptotic expansions. It can sucessfully be extended to physically 
more realistic situations in which coupled sets of truncated Dyson--Schwinger
equations have to be solved allowing for similarly infrared divergent
solutions for the propagators in QCD \cite{Sme97'}.

\section{Description of the program}

\subsection{The main program}

On startup all relevant variables are initialized. These include the maximum
number of iterations (\textit{NITMAX}), the number of meshpoints
(\textit{MESHPTS}) used to represent $\tilde{F}(x)$ on a grid, and the orders
in the asymptotic expansion (\textit{NMAX, MMAX}). Furthermore the infrared
matching point $x_0$ and the ultraviolet cut-off $x_1$ are set to some
appropriate values (\textit{XMIN, XMAX} respectively). The variable
\textit{EPS} is introduced to assess the accuracy of the result. 

After this, the initial gluon function and the corresponding asymptotic
series are generated. In the following, in each iteration a new temporary
gluon function is calculated from equation \eqref{eq:Gluon-DSE+}. In order to
fullfill the constraint \eqref{eq:constraint+} this temporary gluon function
together with the asymptotic series, i.e., the constant $a_{00}$, are rescaled
appropriately before the result of this two--step process is used in the next
iteration. 

Convergence is monitored by pointwise comparing the relative deviation
between the current result for the gluon function and the result of the
previous iteration. Once the maximum deviation is below the desired accuracy
\textit{EPS}, the iteration process is halted. 

Finally, from $\tilde{F}(x)$ the running coupling $\alpha_{\text{S}}(\mu^2) =
g^2 (\mu^2)/4\pi $ is calculated along with the $\beta$-function defined by
$\beta(g)=\mu (dg/d\mu)$. The datafile for the gluon function
(\texttt{gluon.out}) is written in three columns containing (from left) $x =
k^2/\Lambda^2_{\text{MOM}}$,  $\tilde F(x)$, and $\tilde Z(x) = 1/x + \tilde
F(x)$. The running coupling in the format ``$s = \mu^2/\Lambda^2_{\text{MOM}}$,
$\tilde \alpha_{\text{S}}(s)$'',  and ``$g$, $\beta(g)$'' are written to the
datafiles \texttt{alpha.out} and \texttt{beta.out} respectively.

\subsection{Subroutines and functions}

Subroutine \textit{diff} \\
   Calculates the first derivative using a five-point formula
   \cite{Abr84}.

Function \textit{erfc} \\
   Returns the complementary error function $\mathrm{erfc}(x)$
   \cite{Pre94}.

Function \textit{Simpson} \\
   Returns the integral of a function which is given at equally spaced
   abscissas. For a sufficient number of abscissas this function uses
   a closed Simpson rule of order $1/N^4$ \cite{Pre94}.

\section{Testing the program}

Extensive tests were performed to establish resonable ranges to be used for
the infrared matching point $x_0$ als well as the ultraviolet cut-off $x_1$. 
Independence of the results on these matching points was obtained within
these ranges to sufficiently high accuracy. To achieve this, tests have been
performed to find the necessary order in the asymptotic infrared
expansion. The inclusion of even higher orders in this expansion was verified
to have no considerable effect on the results. Similar tests were done to
verify the independence of the (sufficiently large) number of momentum
meshpoints as well as the independence of the initial values chosen for the
gluon renormalization function.

We solved both, Mandelstam's original equation (\ref{eq:Mandelstam}) as well
as the improved equation (\ref{eq:Brown}) obtained by Brown and Pennington in
the same truncation scheme, with the method presented here. Even though both
equations contain rather different coefficients, identical qualitative
features of their solutions were obtained. In particular, the results from
the original Mandelstam equation were verified to reproduce those of
ref.\ \cite{Atk81} where this equation was solved by a different procedure
matching the result of an infrared expansion to a numerically solved
differential equation.   

\section*{Acknowledgments}

We would like to thank F.~Coester for helpful remarks. Most of the work was
done during an appointment of L.v.S.\ at the Physics Division of Argonne
National Laboratory. This work was supported by DFG under contract Al 279/3--1,
by the Graduiertenkolleg T\"ubingen (DFG Mu705/3), by the US Department of
Energy, Nuclear Physics Division, under contract number W-31-109-ENG-38, and
by the BMBF contract number 06-ER-809.

\newpage

\section*{TEST RUN}

\centerline{\textit{standard output}}

\begin{verbatim}
Number of meshpoints :  500
Order of asymptotic series :  M = 4  N = 4
Infrared matching point :  0.20
Ultraviolet cut-off :  1.0E+08
eps =  1.0E-07

Convergence achieved after 136 iterations!
a(0,0) =  2.9446751985E-01
max. deviation between Fin and Fout:  8.95681E-08

\end{verbatim}

\centerline{\textit{gluon.out}}

\begin{verbatim}
    0.4000000000E-09    0.3380217288E-12    0.2500000000E+10
    0.4163493887E-09    0.3556712162E-12    0.2401828914E+10
    0.4333670337E-09    0.3742422549E-12    0.2307512852E+10
                              ...
    0.9230051410E+08    0.6494579660E-01    0.6494580744E-01
    0.9607315655E+08    0.6486822745E-01    0.6486823786E-01
    0.1000000000E+09    0.6479095232E-01    0.6479096232E-01

\end{verbatim}

\centerline{\textit{alpha.out}}

\begin{verbatim}
    0.3641128406E-02    0.9478445822E+06
    0.3789953965E-02    0.8748652799E+06
    0.3944862541E-02    0.8075050070E+06
                              ...
    0.9230051410E+08    0.5300442226E-01
    0.9607315655E+08    0.5287788367E-01
    0.1000000000E+09    0.5275197512E-01

\end{verbatim}

\centerline{\textit{beta.out}}

\begin{verbatim}
    0.6327320355E+02   -0.1244249422E+03
    0.6082727354E+02   -0.1197555950E+03
    0.5847666993E+02   -0.1149508614E+03
                              ...
    0.8161330861E+00   -0.4875758004E-01
    0.8151583181E+00   -0.4857271919E-01
    0.8141872450E+00   -0.4838870479E-01
\end{verbatim}


\begin{thebibliography}{99}
\bibitem{Hau97b}
   A.~Hauck, L.~von Smekal and R.~Alkofer,
   \textit{Solving a Coupled Set of truncated QCD Dyson--Schwinger
   Equations}, submitted to Computer Physics Communications. 
\bibitem{Sme97}
   L.~von Smekal, A.~Hauck and R.~Alkofer, Phys.~Rev.~Lett.\
   \textbf{79} (1997), 3591;
   \newline
   L.~von Smekal, A.~Hauck and R.~Alkofer,
   \textit{A Solution to Coupled Dyson--Schwinger Equations for Gluons and
   Ghosts in Landau Gauge}, hep--ph 9707327, e--print, submitted to
   Ann.~Phys.
\end{thebibliography}

\begin{thebibliography}{99}
\bibitem{Man79}
   S.~Mandelstam, Phys.~Rev.~D \textbf{20} (1979), 3223.
\bibitem{Atk81}
   D.~Atkinson et al., J.~Math.~Phys.\ \textbf{22} (1981), 2704;
   \newline
   D.~Atkinson, P.~W.~Johnson and K.~Stam, J.~Math.~Phys.~\textbf{23} (1982),
   1917.
\bibitem{Bro89}
   N.~Brown and M.~R.~Pennington, Phys.~Rev.~D \textbf{39} (1989), 2723;
   \newline
   N.~Brown, Ph.~D.\ Thesis, University of Durham, August 1988.
\bibitem{Sme97'}
   L.~von Smekal, A.~Hauck and R.~Alkofer, Phys.~Rev.~Lett.\
   \textbf{79} (1997), 3591;
   \newline
   L.~von Smekal, A.~Hauck and R.~Alkofer,
   \textit{A Solution to Coupled Dyson--Schwinger Equations for Gluons and
   Ghosts in Landau Gauge}, e--print hep--ph 9707327, submitted to Ann.~Phys.
\bibitem{Atk97}
   D.~Atkinson and J.~C.~R.~Bloch, e--print hep--ph/9712459.
\bibitem{Hau97'}
   A.~Hauck, L.~von Smekal and R.~Alkofer,
   \textit{Solving a Coupled Set of truncated QCD Dyson--Schwinger
   Equations}, submitted to Computer Physics Communications. 
\bibitem{Bro88}
   N.~Brown and M.~R.~Pennington,
   Phys.~Rev.~D \textbf{38} (1988), 2266.
\bibitem{Bal80}
   J. S. Ball and T.-W. Chiu, Phys.~Rev.~D \textbf{22} (1980), 2550;
   \newline
   S.~K.~Kim and M.~Baker, Nucl.~Phys.\ \textbf{B164} (1980), 152.
\bibitem{BBZ81}
   M.~Baker, J.~S.~Ball and F.~Zachariasen, Nucl.~Phys.\ \textbf{B186} (1981),
   531; 560.
\bibitem{Scho82}
   W.~J.~Schoenmaker, Nucl. Phys. \textbf{B194} (1982), 535.
\bibitem{CuRo91}
   J.~R.~Cudell and D.~A.~Ross, Nucl.~Phys.\ \textbf{B358} (1991), 247. 
\bibitem{Bue95}
   K.~B\"uttner and M.~R.~Pennington, Phys.~Rev.~D \textbf{52} (1995), 5220. 
\bibitem{MPpriv}
   R.~Alkofer, M.~R.~Pennington, L.~von Smekal and P.~Watson,
   work in progress.
\bibitem{Len94}
   F.~Lenz, H.~W.~L.~Naus, K.~Otha and M.~Thies, Ann.~Phys.\
   \textbf{233} (1994), 12; 51;
   \newline
   F.~Lenz, H.~W.~L.~Naus and M.~Thies, Ann.~Phys.\ \textbf{233} (1994), 317.
\bibitem{sme91}
   L.~v.~Smekal, P.~A.~Amundsen and R.~Alkofer, Nucl.~Phys.\ \textbf{A529}
   (1991), 633.
\bibitem{PDG96}
   R.~M.~Barnett et al., Phys.~Rev.~D \textbf{54} (1996), sect.~9, p.~77-84. 
\bibitem{Eic80}
   E.~Eichten, K.~Gottfried, T.~Kinoshita, K.~D.~Lane and T.~M.~Yan,
   Phys.~Rev.~D \textbf{21} (1980), 203.
\bibitem{Buc81}
   W.~Buchm\"uller and S.~H.~H.~Tye, Phys.~Rev.~D \textbf{24} (1981), 132.
\bibitem{Din90}
   H.~Q.~Ding, C.~F.~Baillie and G.~C.~Fox, Phys.~Rev.~D \textbf{41}
   (1990), 2912.
\bibitem{Abr84}
   M.~Abramowitz and I.~A.~Stegun, eds.,
   \textit{Pocketbook of Mathematical Functions}
   (Verlag Harri Deutsch, Frankfurt/Main, 1984).
\bibitem{Pre94}
   W.~H.~Press, S.~A.~Teukolsky, W.~T.~Vetterling and B.~B.~Flannery,
   \textit{Numerical Recipes in FORTRAN}
   (Cambridge Univ.~Press, Cambridge, 1994).
\end{thebibliography}
\end{document}